# Direction-dependent Optical Modes in Nanoscale Silicon Waveguides

Jacob T. Robinson[1†] and Michal Lipson[1,2]*

[1]*School of Electrical and Computer Engineering, Cornell University, Ithaca, NY 14853*
[2]*Kavli Institute at Cornell for Nanoscale Science, Cornell University, Ithaca, NY 14853*
[†]*Current address: Department of Chemistry and Chemical Biology, Harvard University, Cambridge, MA 02138*
*Corresponding author: ml292@cornell.edu

On-chip photonic networks have the potential to transmit and route information more efficiently than electronic circuits[1,2]. Recently, a number of silicon-based optical devices including modulators[3-5], buffers[6], and wavelength converts[7,8] have been reported. However, a number of technical challenges need to be overcome before these devices can be combined into network-level architectures. In particular, due to the high refractive index contrast between the core and cladding of semiconductor waveguides, nanoscale defects along the waveguide often scatter light into the backward-propagating mode[9]. These reflections could result in unwanted feedback to optical sources[10] or crosstalk in bi-directional interconnects such as those employed in fiber-optic networks[11]. It is often assumed that these reflected waves spatially overlap the forward-propagating waves making it difficult to implement optical circulators or isolators which separate or attenuate light based on its propagation direction[12-15]. Here, we individually identify and map the near-field mode profiles of forward-propagating and reflected light in a single-mode silicon waveguide using transmission-based near-field scanning optical microscopy (TraNSOM) [16,17]. We show that unlike fiber-optic waveguides, the high-index-contrast and nanoscale dimensions of semiconductor waveguides create counter propagating waves with distinct spatial near-field profiles. These near-field differences are a previously-unobserved consequence of nanoscale light confinement and could provide a basis for novel elements to filter forward-propagating from reflected light.

The high-refractive-index contrast and nanoscale dimensions of semiconductor waveguides not only enable dense device integration, but can also fundamentally affect the properties of optical propagation[2]. The behavior of forward-propagating and reflected modes is particularly important for developing strategies to overcome potentially harmful consequences of reflection[10,11]. While fiber-optic interconnects are often useful models for nanophotonic waveguides, they fail to reproduce the effects of nanoscale light confinement on the near-field properties of counter-propagating waves. In fiber-optic interconnects, forward-propagating and reflected waves spatially overlap in the absence of non-reciprocal components such as magnet-optic materials. By design, the dimensions of single mode fiber-optic waveguides allow only one mode to propagate for each orthogonal polarization[10]. Since the fibers are often radially symmetric (and the difference in refractive index is not large enough to cause significant electric field discontinuities) the orthogonally polarized modes are typically degenerate and their intensity profiles are nearly identical. Propagating light is therefore confined to this single mode and the distribution of optical intensity for forward-propagating and reflected light is identical in the near-field. Even when the degeneracy between orthogonally polarized modes is broken (by introducing asymmetry in the fiber cross-section[18]), reflected light typically remains in the polarization state of the forward-propagating mode, and is therefore indistinguishable based on its near-field intensity distribution. This spatial indistinguishability makes it difficult to separate forward-propagating from reflected light.

Here, using near-field microscopy, we show that in nanophotonic waveguides forward-propagating and reflected waves are spatially distinct. Figure 1 shows a schematic of the experiment. Optical fibers are used to couple light into and out of the waveguide (shown in green). The waveguide is fabricated in silicon on insulator. Details of the experimental setup and device fabrication can be found in the Methods section. While scanning the waveguide with an Atomic Force Microscope (AFM) probe we constantly monitor the power transmitted through the device. This technique, Transmission-based Near-field Scanning Optical Microscopy (TraNSOM), was recently developed for imaging near field profiles in high index-contrast-waveguides [16,17] and subsequently applied to optical resonant cavities[19-21]. While phase-sensitive[22,23] and time-resolved[24] near-field microscopy techniques can determine the direction of optical propagation based on the sign of the propagation constant, here we use TraNSOM to search for spatial differences in intensity profiles between forward-propagating and reflected modes. When the probe interacts with the evanescent field of the guided wave it scatters some of the light out of this mode. Since the probe is in the near field of the waveguide, much of this scattered light couples back into the guided mode and propagates in a direction opposite to the incident light[9]. This is measured as a probe-induced reflection and can be used to determine the propagation direction of the incident light. For instance, when the probe interacts with the forward-propagating mode, light is reflected away from the output and the power transmitted decreases. Conversely, when the probe interacts with the backward-propagating reflected light, probe-induced reflection redirects some of this light toward the output. Thus probe interaction with the reflected light results in an increase in the optical power detected at the output. By looking for transmission changes of opposite sign, we aim to distinguish forward-propagating from reflected light. If the mode profiles of the forward-propagating and reflected waves are identical (as is expected for low-index waveguides like fiber optics) we should observe that scattering by the probe only decreases the transmitted power. This is because at every point across the waveguide, the probe would interact simultaneously with both forward-propagating and reflected light. Due to propagation losses, the amplitude of the forward propagating mode is larger and would therefore dominate the measured signal. If, however, the forward-propagating and reflected waves are spatially separated (i.e. at specific points across the waveguide the probe interacts with one wave and not the other), we should be able to observe *both* a decrease *and* increase in transmission as the probe interacts individually with either the forward-propagating or reflected waves respectively.

Figure 2a shows the result of this measurement where both forward-propagating and reflected waves are distinct and clearly visible. The measured topography of the waveguide is shown in the Fig. 2a insert, along with the simultaneously measured transmitted power (Fig. 2a). The measurement is performed with a source wavelength of 1.532 µm (see Methods). Dark blue regions (point **A**) indicate a probe-induced decrease in transmitted power resulting from interaction with forward-propagating light. Red regions (point **B**) indicate a probe-induced *increase* in transmitted power resulting from interaction with *backward*-propagating reflected light. Thus the sign of the transmission change indicates the direction of light propagation. (Note the small changes in transmission when the probe is far from the waveguide are the result of far-field suppression and enhancement of radiation scattered from defects along the waveguide[25]).

We verify that backward propagating light is responsible for the measured increase in transmission by eliminating its contribution to the measured signal and repeating the TraNSOM measurement. This is achieved by using an optical source with a short coherence length. Figure 2b shows the TraNSOM image using an optical source with a 1.4 mm coherence length. Because this coherence length is shorter than the path from the probe to the end of the waveguide and back, by the time the reflected light is scattered by the probe it has no well-defined phase relationship with the forward-propagating light. Therefore the scattered light from the backward-propagating reflected wave is equally likely to constructively or destructively interfere with the forward-propagating mode and thus has no net effect on the transmitted power. Therefore, by using a short-coherence-length source we can selectively map the distribution of only the forward-propagating light. As expected, Fig. 2b shows no probe-induced increases in transmission. This verifies that the measured increase in transmission is the result of interaction with the backward-propagating reflected light in the guided mode. Figure 2c shows a theoretical TraNSOM signal calculated with contributions from both forward-propagating and reflected light. Figure 2d shows the same calculation considering only forward-propagating light. This corresponds to the measured data in Fig. 2a and b respectively. Details of the calculation are described in subsequent sections.

The spatially-distinct near-field profiles observed in Fig. 2 are the result of large minor field components in nanoscale high-index-contrast waveguides. Figure 3a and b show the calculated vertical ($y$) component of the electric field for the fundamental TM and TE modes respectively. We see in Fig. 3b that although the TE mode is polarized primarily in the horizontal ($x$) direction there are large minor field components in the $y$-direction at the waveguide corners. Due to the large minor field components these modes are frequently referred to as quasi-TE or quasi-TM. If both TE and TM modes are excited the total shape of the mode profile is the coherent sum of these two modes. This causes the total mode profile to "lean" to the left or right depending on the phase difference between the TM and TE modes (see Fig. 3c and d) [26]. When the forward propagating wave encounters a perturbation to the waveguide width, such as the narrow tapers used to couple light on and off chip[27] (see Figs. 1 and S1), the reflected TE mode acquires a $\pi$ phase shift, while the TM mode does not[28]. This causes the backward-propagating wave to

"lean" in the opposite direction as the forward-propagating light. This property of large minor field components at the waveguide corners is unique to high-index nanoscale waveguides. In fiber optic waveguides, the index contrast is too small to produce these field components, and thus the shape of the mode profile does not depend on the phase between the TE and TM modes.

Simulations using a 3D Finite Difference Time Domain (FDTD) method verify that forward and backward propagating waves "lean" in opposite directions. Since rectangular high-index-contrast waveguides are typically highly birefringent, the TE and TM modes propagate with different phase velocities due to their different effective indices [29]. The evolving phase difference between the TE and TM modes causes the near-field intensity distribution to oscillate between left and right "leaning" profiles. This can be seen in 3D-FDTD simulations where both the TE and TM modes were launched from left to right (Fig. 3e). Here we plot an *x-z* cross section through the waveguide 200 nm above the silicon oxide substrate. The period of oscillation (*L*) for this experiment is determined by the birefringence ($\Delta n_{eff}$=0.46 from finite element mode solver) and the wavelength ($\lambda$=1.532 µm): $L = \lambda / \Delta n_{eff} = 3.3$ µm. This period matches the beat period measured by TraNSOM in Fig. 2a and b. The beating observed here is similar to the polarization mode beating observed in birefringent fibers[18]; however, in low-index-contrast fibers, due to the negligible minor field components, the shape of the mode profile does not change as it propagates. The phase of the oscillation is determined by the initial polarization state (phase relationship between the TE and TM components). The backward-propagating reflected wave is simulated by adding a π phase shift to the TE mode and launching both TE and TM modes from the right of the waveguide (Fig. 3f). As expected, the backward-propagating and forward-propagating light oscillate out of phase with one another, i.e. at each point in the waveguide the modes "lean" in opposite directions. For example, the forward-propagating mode leans toward the point labeled **A'** (Fig. 3e) while the backward propagating mode leans away (Fig. 3f). Point **B'** shows the opposite behavior. These points correspond to points **A** and **B** in Fig. 2. One should note that this behavior does not violate time-reversal symmetry since switching the input and output ends of the waveguide results in identical behavior.

The measured data in Fig. 2 matches analytical models which incorporate effects for both forward and backward propagating modes. We can write the change in total power collected at the waveguide output (as a function of probe location $x$) in terms of the amount of light scattered by the probe:

$$\frac{\Delta T}{T_0}(x) = -\frac{Q}{Z_0}\int_{A(x)} \left| a_{TM} E_{TMy} e^{ik_{TM}z} + a_{TE} E_{TEy} e^{ik_{TE}z} \right|^2 dA + \eta \frac{Q}{Z_0}\int_{A(x)} \left| b_{TM} E_{TMy} e^{ik_{TM}z} - b_{TE} E_{TEy} e^{ik_{TE}z} \right|^2 dA. \quad (1)$$

The two terms to the right of the equality represent the scattering of forward-propagating and reflected light respectively (note the sign difference). Also, note the phase (sign) change added to the TE mode in the second term, which is the result of reflection. Here the *TM* and *TE* subscripts denote the TE or TM mode respectively and $E_y$ is the *y*-component of the electric field. The cross-sectional profile of the probe is written as *A* (see Methods), $x$ and $z$ are Cartesian coordinates, $k$ is the propagation constant, and $Z_0$ is the free space impedance. The scattering efficiency ($Q$) is the measured to be ~25 (see Supplementary Information (SI) and Fig. S2). Due to the large index contrast between the probe and air-cladding, the efficiency with which scattered light couples to the counter-propagating mode ($\eta$) is expected to be near be unity [9]. For simplicity, we will take this factor to be 1. The amplitudes of the forward and backward propagating modes are written as *a* and *b* respectively. We can write *b* in terms of *a* according to:

$$b_{TM,TE} = a_{TM,TE} e^{-\alpha_{TM,TE} 2l} R_{TM,TE} \quad , (2)$$

where α is the waveguide loss per unit length, *l* is the distance between the probe and the source of reflection (the waveguide output in this case) and *R* is the reflectivity. We take the waveguide loss to be -6.64 dB/cm and -15.36 dB/cm for the TM and TE modes respectively. These values are taken from similar waveguides fabricated in silicon (see Methods) and measured using the cut-back method [30]. The coupling efficiency of each polarization and their respective reflection at the waveguide interfaces is difficult to measure directly; however, the reflection is known to be high, particularly for the TM mode, which despite having lower propagation loss, transmits -11 dB less power through the device as compared to the TE mode.

Based on our analytical model, we identify specific values of reflectivity and ratios of TE to TM excitation which allow forward-propagating and reflected light to be distinguished. Figure 4 shows the maximum change in transmission ($\max[\Delta T/T_0]$) as a function of reflectivity at the chip interface ($R$) and the relative amplitude of TE mode ($|a_{TE}|^2$). This is calculated using equation (1) and (2), where the probe position ($x$) is fixed at the location of maximum modal overlap. We keep constant the amplitude of the TM mode ($|a_{TM}|^2 = 1$, since $\max[\Delta T/T_0]$ depends only the ratio of the modal amplitudes). The reflectivity of the TE mode ($R_{TE}$) is set to %10 of $R_{TM}$ based on the relative output powers and propagation losses measured above. When scattering by the probe results only in a decrease in power transmitted through the waveguide, $\max[\Delta T/T_0] = 0$ and the forward-propagating and reflected propagating modes cannot be disambiguated. This occurs when the absolute value of the first term in equation (1) is always larger than that of the second term. In other words, at each probe position more light is scattered from the forward-propagating light than from the reflected light. We refer to this as the "normal" scattering regime, since as expected, introduction of a scattering point results in a decrease in power transmitted through the waveguide. Conversely, where $\max[\Delta T/T_0] > 0$, forward-propagating and reflected light can be distinguished. In this regime, a scattering point can redirect the backward-propagating reflected light such that the power transmitted through the waveguide increases. Since this is an unexpected consequence of near-field scattering, we refer to this as the "anomalous" scattering regime. The waveguide used for these experiments was designed with $R_{TE} \approx 0.75$ such that the anomalous regime is accessible for ratios of TE to TM excitation > 0.5 which is easily achieved using a polarization controller.

We verify our measured results by reproducing the data in Fig. 2a and b using our analytical model. We select from Fig. 4, the point $R_{TM} = 0.75$, and $|a_{TE}|^2 = 0.8$ since this best matches max[$\Delta T/T_0$] measured in Fig. 2a. According to equations (1) and (2) we can calculate $\Delta T/T_0$ as a function of $x$ and $z$. The result plotted in Fig. 2c shows excellent agreement between our model and the measured data shown in Fig. 2a. To confirm that the positive values correspond to backward-propagating reflected light we can remove the second term from equation (1) to consider only the forward-propagating mode. Recalculating the image (Fig. 2d) shows that as expected, in the absence of the backward-propagating mode, probe-induced scattering only decreases the power transmitted through the waveguide. This is consistent with our measurement in Fig. 2b, in which we experimentally isolate the effect of only the forward-propagating light. This agreement between our model and the measured data confirms that unlike micron-scale low-index fiber optics, nanoscale waveguides can posses forward and backward traveling waves with unique near-field profiles.

The distinct near-field profiles for counter-propagating waves reveal fundamental differences between optical propagation in nanoscale waveguides compared to free-space and fiber optics. While this phenomenon is dependent on input polarization and reflectivity of the TE and TM modes, it is solely a consequence of strong optical confinement and is likely to occur in the nanoscale high-index waveguides that are used widely in industrial and academic research labs. In addition to potentially affecting device performance, this phenomenon could be utilized as a basis for selectively attenuating reflected waves. Active components or specific waveguide geometries could be developed to create uni-directional waveguides which could limit the intensity of reflected light. This would provide a path toward developing robust nanophotonic devices and architectures unhindered by optical reflections.

**Methods:**

**Fabrication:** Silicon-on-insulator (SOI) waveguides are approximately 1 cm long, 460 nm wide, and 260 nm tall covered with a 130 nm tall thermally-grown oxide which served as a hard mask during reactive ion etching. SOI substrates purchased from Soitec have a 3 μm thick buried oxide layer. The fabrication method is similar to that described in ref. 7.

**Simulation:** The cross-sectional profile $A$ is an inverted triangle with a half angle of $10^o$ at the apex (Manufacturer's specifications, Nanosensors). At each position ($x$) the $y$ coordinate of the profile is chosen such that the probe is positioned in contact with the surface (see Supporting Information (SI)). Finite difference mode solvers and 3D-FDTD code was developed by Christina Manolatou. Analysis and figures were generated using Matlab.

**Measurement:** The optical source is a multi-line external cavity laser (ECL) amplified with a 120 mW EDFA (JDS Uniphase) and filtered using a 1.4 nm FWHM tunable bandpass filter (TBF) centered at 1532 nm. The linewidth of the external cavity laser was less than the resolution of our optical spectrum analyzer (<10 pm). The output of the TBF was sent through a digital polarization controller (HP) which we used to polarize the input to excite a combination of TE and TM modes. For the short-coherence-length measurements the ECL was turned off leaving 1.4 nm bandwidth amplified spontaneous emission as our source. Light was coupled into and out of the waveguide using fibers glued to the waveguide facets using a UV curable epoxy (see ref. 7 for details of the packaging). In all cases, a single optical source is used to illuminate the waveguide from only the input side. To facilitate optical coupling, the waveguides widths were tapered (similar to ref. 19) from 460 nm to 120 nm linearly over a length of 100 μm and clad with Shipley 1818 photoresist (see Fig. S1). The waveguide was imaged with a Dimension 3100 atomic force microscope using a Pt/Ir coated AFM probe from Nanosensors. The waveguide output was collected into a fiber glued to the waveguide facet and measured with a Newport 1818-IG photodetector and 2832c power meter. The analog output of the power meter was amplified using an HP voltage pre-amplifier with a 30 Hz low pass filter and then sent to the analog input of the AFM for simultaneous recording with the waveguide topography.

**Acknowledgments:** The authors thank A. Falk, F. Koppens, and N. Snapp for their comments on the manuscript. Research support is gratefully acknowledged from the National Science Foundation's CAREER Grant No. 0446571. This work was performed in part at the Cornell NanoScale Facility, a member of the National Nanotechnology Infrastructure Network, which is supported by the National Science Foundation (Grant ECS 03-35765) and we made use of STC shared experimental facilities supported by the National Science Foundation under Agreement No. ECS-9876771.

**Figures:**

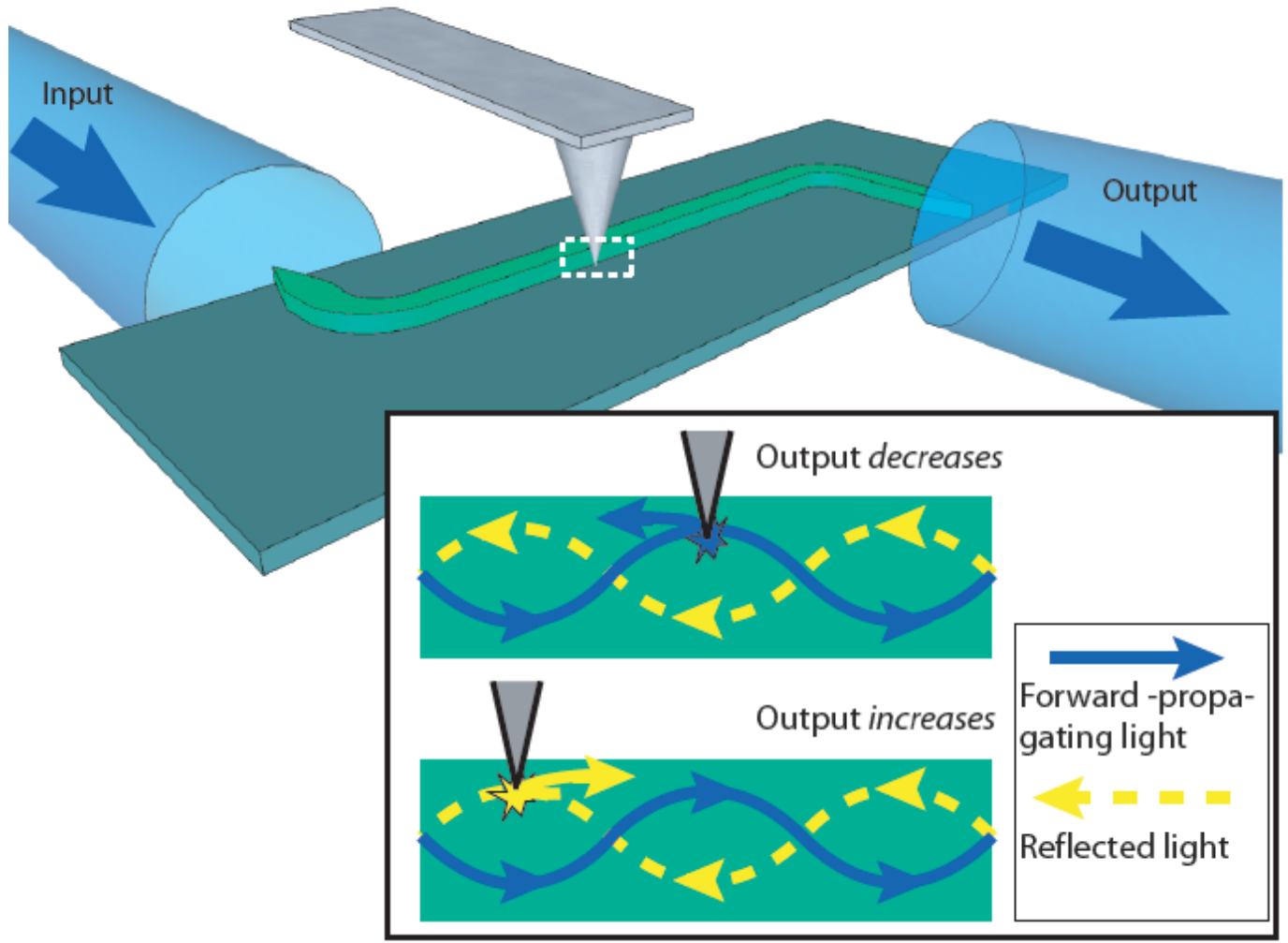

**Figure 1| Schematic of TraNSOM measurement of counter-propagating mode profiles.**

Fiber optics couple light into and out of the waveguide. The power transmitted through the device is constantly monitored as the waveguide is scanned by an AFM probe. Probe-induced scattering of the forward-propagating or reflected light decrease or increase the output power respectively (inset: solid and dashed lines respectively). Based on the sign of the change in transmitted power, forward-propagating and backward-propagating reflected light can be distinguished.

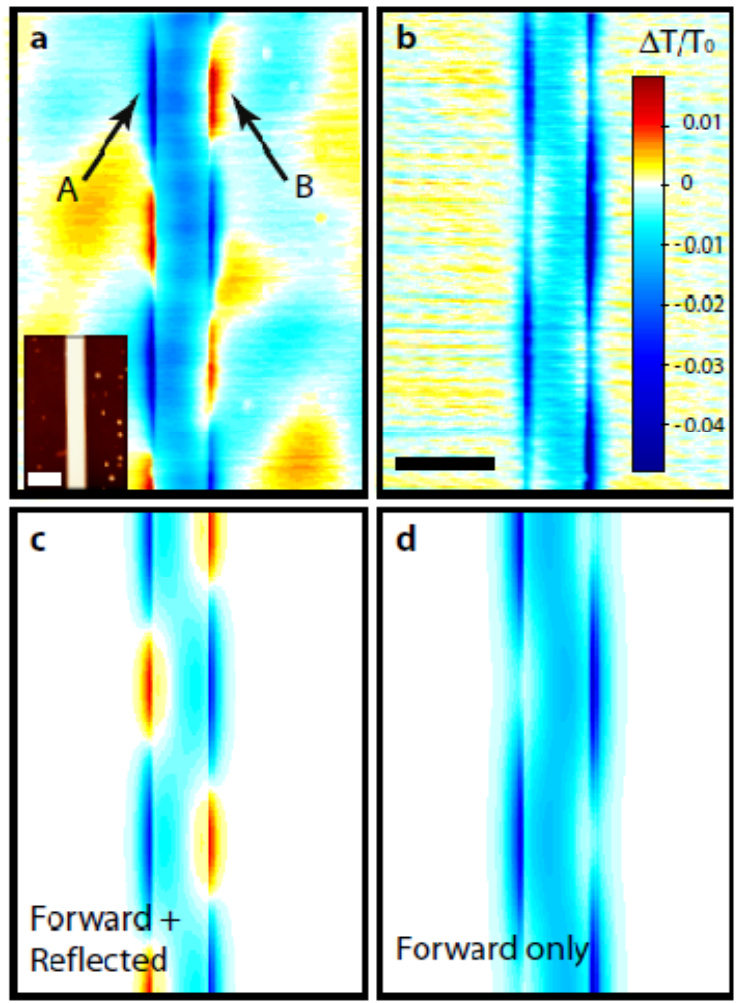

**Figure 2 | Measurement and theory of counter-propagating waveguide modes. a,** measured change in transmission as a function of probe position for a 5 μm length of SOI waveguide. Point **A** corresponds to a region where the transmission *decreases* indicating light here is traveling in the *forward* direction. Point **B** corresponds to a region where the transmission *increases* indicating light here is traveling in the *backward* direction. **a inset,** simultaneously measured AFM topography. **b,** change in transmission vs. probe position for the same length of waveguide using a short-coherence-length source which isolates the contribution of forward-propagating mode. The color scale in (**b**) for the relative change in transmission is the same for (**a**)-(**d**). **c,**

calculated near-field image of the probe-induced change in transmission over a 5 µm segment of waveguide according to our model including both forward-propagating and reflected light. This corresponds to the measured data in (**a**). **d,** calculated near-field image considering only forward propagating light corresponding to (**b**). Scale bars are 1 µm.

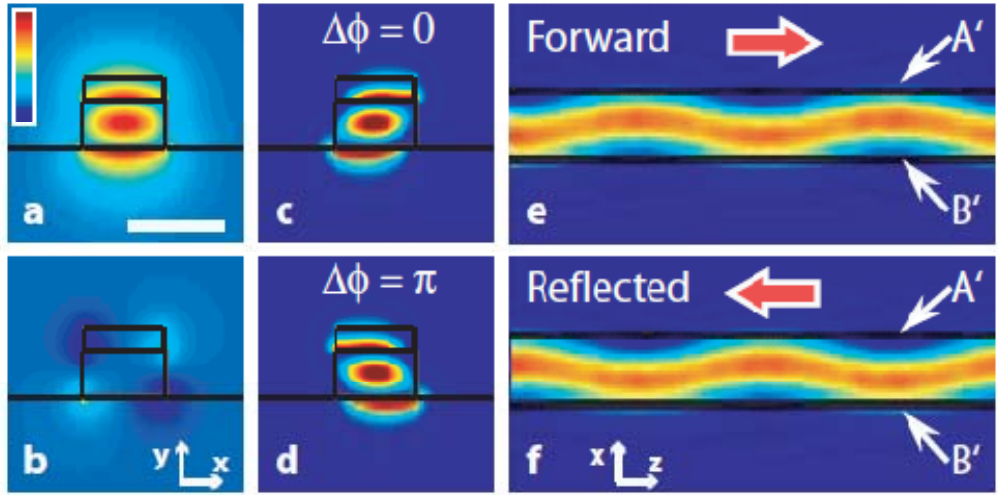

**Figure 3 | Interaction between TE and TM modes in high index contrast waveguides. a** and **b,** calculated *x-y* cross-sectional mode profiles for *y* component of electric field ($E_y$) of the TM and TE modes respectively plotted on the same color scale. Note that although the *y* component of the TE mode (**b**) is not the major field component, due to the nanoscale waveguide geometry, it is only slightly smaller in magnitude compared to the major component of the TM mode (**a**). This causes the orthogonally polarized modes to interact. **c** and **d**, $|E_y|^2$ for the TM and TE modes summed in-phase and out-of-phase respectively. In (**a**)-(**d**) the TE mode power has been multiplied by a factor of 4 relative to the TM mode. **e** and **f,** *x-z* cross sections of 3D-FDTD simulations of the evolution of an optical mode consisting of both TE and TM components as it propagates in the forward (*+z*) and backward (*-z*) directions respectively. The forward-propagating and reflected light "lean" toward the points labeled **A'** and **B'** respectively. These points correspond to **A** and **B** in Fig. 2. Scale bar in (**a**) is 1 μm. All figures are plotted at the same scale.

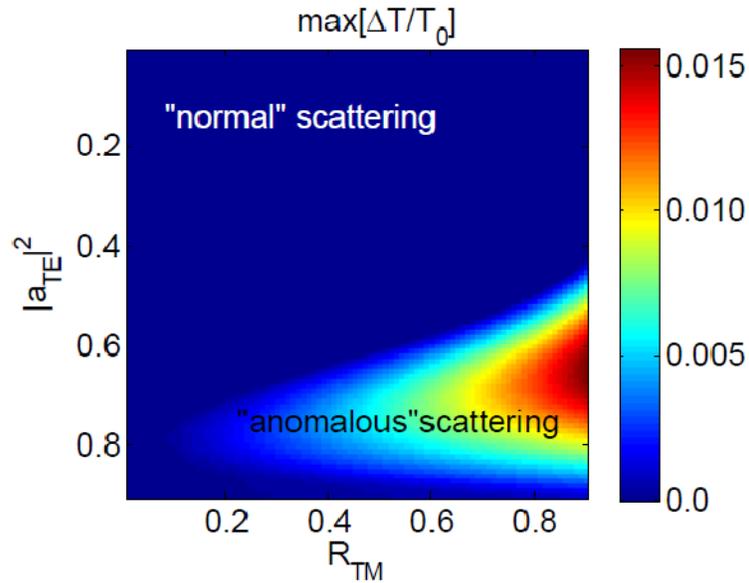

**Figure 4 | Phase diagram for differentiating propagation direction.** Maximum probe-induced change in transmission calculated according to Eq. 1 as a function of relative TE mode amplitude squared ($|a_{TE}|^2$) normalized to total power, and TM reflectivity at the waveguide-fiber interface. The relative reflectivity of the TE mode ($R_{TE}$) is fixed at $0.1*R_{TM}$. The "anomalous" scattering regime refers to the region where scattering by the probe results in an increase in the amount of power transmitted through the waveguide. In this region forward-propagating and reflected light can be distinguished by near-field scattering.